\documentclass[%
 aps,            
 prl,            
 reprint,       
 floatfix,      
 amsmath,amssymb 
]{revtex4-2}
\usepackage{graphicx}
\usepackage{dcolumn}
\usepackage{bm}
\usepackage{hyperref}
\usepackage{color}
\usepackage{braket}

\begin{document}

\preprint{APS/123-QED}

\title{Long-distance distribution of atom-photon entanglement based on a cavity-free cold atomic ensemble }

\author{Tian-Yu Wang$^{1,2,3,\ddag}$, Ren-Hui Chen$^{1,2,3,\ddag}$, Yan Li$^{1,2,\ddag}$, Ze-Hao Shen$^{4,\ddag}$, Xiao-Song Fan$^{1,2,3}$, Zheng-Bang Ju$^{1,2,3}$, Tian-Ci Tang$^{1,2,3}$, Xia-Wei Li$^{1,2,3}$, Jing-Yuan Peng$^{1,2,3}$, Zhi-Yuan Zhou$^{1,2,3,*}$, Wei Zhang$^{3,\dag}$, Guang-Can Guo$^{1,2,3}$, Bao-Sen Shi$^{1,2,3,\S}$}
\affiliation{
$^1$Laboratory of Quantum Information, University of Science and Technology of China, Hefei 230026, China\\
$^2$Anhui Province Key Laboratory of Quantum Network,
University of Science and Technology of China, Hefei 230026, China\\
$^3$Hefei National Laboratory, University of Science and Technology of China, Hefei 230088, China\\
$^4$School of Physics, Xi’an Jiaotong University, Xi’an 710049, China
}
\altaffiliation{$^*$zyzhouphy@ustc.edu.cn}

\altaffiliation{\\$^\dag$changong@ustc.edu.cn}

\altaffiliation{\\$^\S$drshi@ustc.edu.cn}

\altaffiliation{\\$^\ddag$These authors contributed equally to this work }

\date{\today}

\begin{abstract}

Constructing a quantum memory node with the ability of long-distance atom-photon distribution is the essential task for future quantum networks,
enabling distributed quantum computing, quantum cryptography and remote sensing. Here we report the demonstration of a quantum-network node with a simple cavity-free cold atomic ensemble. 
This node gives an initial retrieval efficiency of approximately 55\% and memory lifetime of 160 $\mu$s for atomic qubits. With the aid of a high-efficiency and polarization-independent quantum frequency conversion (QFC) module, the generated entangled photon in the node at 780-nm wavelength is converted to telecom S band at 1522 nm, enabling atom-photon distribution over long distance.
We observe an entanglement fidelity between the atoms and telecom photon exceeding  80\% after photon transmission over 20-km fiber {with an end excitation probability of 0.2\% and repetition of 1.7 kHz}, the remaining infidelity being dominated by atomic decoherence.
The low-noise QFC with an external efficiency up to 48.5\% gives a signal-to-noise-ratio of 6.9 for transmitted photons with fiber length up to 100 km, laying the cornerstone for entanglement distribution at a hundred-km level. This result provides a new platform towards the realization of a long-distance quantum network.

\end{abstract}

\maketitle
\section{Introduction}
Quantum repeaters \cite{Zoller-1998-PRL}, allowing for long-distance quantum communication with a distance that far beyond direct transmission can reach, are very useful for distributed quantum computing \cite{Kim-2014-PRA,Chou-2018-Nature}, quantum cryptography \cite{Gisin-2002-RMP} and quantum sensing \cite{Gottesman-2012-PRL,Lukin-2014-Nat.Phys}. In such a repeater-based network \cite{Kimble-2008-Nature}, a key ingredient is the quantum node, which can generate entanglement and share it with a distance. So far, various systems are proved capable to serve as the quantum node, for example, atomic ensembles \cite{Bao-2022-PRL,Bao-2024-Nature}, single atoms \cite{Gerhard-2012-Nature, Gerhard-2021-Science, Zhang-2022-Nature}, trapped ions \cite{Blinov-2004-Nature, Hucul-2015-Nat.Phys.,Nadlinger-2022-Nature,Lanyon-2023-PRL,Pu-2025-PRL}, defect systems \cite{Hanson-2015-Nature,Lukin-2024-Nature} and solid systems \cite{de.Riedmatten-2021-Nature, Zhou-2021-Nature} and so on.

For an ideal quantum node, there are usually two requirements which must be met. For one, the node itself has to be robust, demanding a long coherence time and a high readout efficiency \cite{Sangouard-2011-Rev.Mod.Phys.}, which means the coherence time should be at least longer than the single-trip photon transmission time and memory efficiency should be high enough, usually above 50\% to beat the no-cloning limit \cite{no-clone-limit-2001-PRA}. For another, the node should be able to generate entangled photons at telecom wavelength \cite{zhang-2016-pra,pu-2019-prx}or equipped with a high-efficient quantum frequency conversion (QFC) module \cite{Kennedy-2010-Nat.Phys., De.Greve-2012-Nature, de.Riedmatten-2017-Nature, Ikuta-2018-Nat.Commun., Jürgen-2018-Nat.Commun., Hanson-2019-PRL, Lanyon-2019-npjQuantumInf., Weinfurter-2020-PRL, Bao-2022-PRL} to guarantee a low-loss photon transmission over long fibers. Such systems have recently been developed to demonstrate the longest atom–photon distribution distance—over 101-km fiber in both single-atom \cite{Weinfurter-2024-PRX} and three-ion \cite{Krutyanskiy-2024-PRX} systems, where atomic readout is implemented after single-trip time with unitary efficiency.

Among all quantum-node platforms, the atomic-ensemble system with proved features such as subsecond lifetime, high retrieval efficiency \cite{Bao-2016-NaturePhotonics} and multiplexing capability \cite{Ding-2016-LightSci.Appl, Pu-2024-NatureCommunications, Wang-2025-Optica}, is widely regarded as one of the most promising candidates. 
Recent demonstrations including postselected entanglement between two atomic ensembles separated by 12.5 km \cite{Bao-2022-PRL} and three-node network with each fiber at 10-km scale \cite{Bao-2024-Nature}, where single-trip time or even round-trip time for readout delay is implemented.
However, all recent advances are accomplished with ring-cavity schemes, raising concerns about its scalability when considering the increased complexity and difficulty.

Here we report the demonstration of a quantum-network node with a simple cavity-free cold atomic ensemble. 
This ensemble gives an initial retrieval efficiency of approximately {55}\%, a record that has never been reported in such {a cavity-free cold atomic ensemble system by using an improved Duan-Lukin-Cirac-Zoller (DLCZ) protocol \cite{DLCZ-2001-Nature}}, and lifetime of 160 $\mu$s with the increased optical depth (OD) to 10 and precise control of magnetic field.
First, the entanglement between polarization state of a 780-nm photon and atomic spin wave in the node is probabilistically generated. The photon is then coupled into a QFC module and converted to telecom S band at 1522 nm and later transferred over long optical fiber spools, where the polarization drift is automatically compensated. Thanks to the high-efficiency QFC and robust memory, the entanglement can still be verified after 20-km transmission when a single-trip readout delay is implemented. The low-noise QFC with an external efficiency up to 48.5\% gives a signal-to-noise-ratio of 6.9 for transmitted photons with fiber length up to 100 km, laying the cornerstone for entanglement distribution at a hundred-km level.
Based on our results, we analyze the potential improvement for long-distance quantum network links in such systems.

\begin{figure*}[t]
\centering
\includegraphics[width=\textwidth]{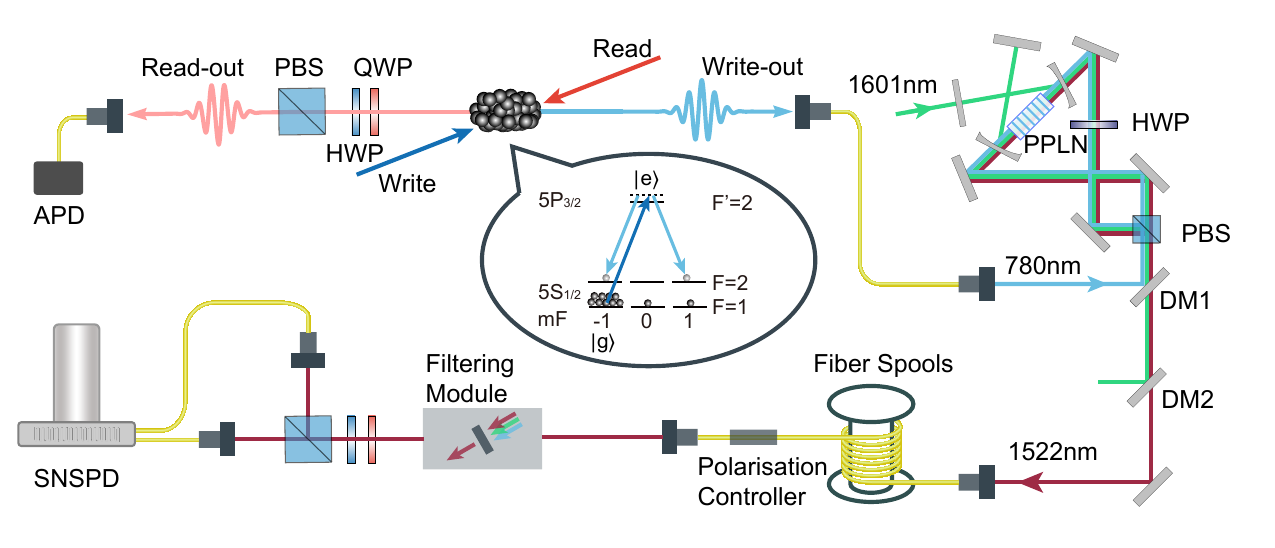}
\caption{Simplified experimental layout. The $^{87}\mathrm{Rb}$ atomic-ensemble module employs a counter-propagating write-read geometry whose beams have a waist of $\sim$240 $\mu$m, the write-out and read-out modes adopt the same layout but with a reduced beam diameter of $\sim$70 $\mu$m. The two optical axes are separated by 0.9$^\circ$. 
After SRS, the entangled write-out photon at 780-nm wavelength is coupled into a QFC module where it is mixed with 1601-nm pump light in a bulk periodically poled lithium niobat (PPLN) crystal within a Sagnac interferometer, and converted into 1522-mm photons for long fiber transmission up to 20 km. 
The polarization drift in long fibers is automatically compensated 
{see more details in {Supplemental Material \cite{SupplementalMaterial}} part 5}.
After long fiber transmission, the telecom single photon is first conducted to a filtering module 
and then projected on a PBS and detected by SNSPDs. The successful detection of 1522-nm photons triggers a read process on atoms, later on the emitted read-out single photon is projected on a PBS and detected by an APD. Here, the atomic ensemble and the QFC module with SNSPDs are housed in separate rooms, a communicating system are used, {see more details in {Supplemental Material \cite{SupplementalMaterial}} part 6}. 
INSET: The energy scheme to generate atom-photon entanglement.
PBS: Polarizing Beam Splitter, QWP: Quarter-Wave Plate, HWP: Half-Wave Plate, APD: Avalanche Photodiode, DM: dichroic mirror. 
}
\label{Fig1}
\end{figure*}

\section{Methods}

The experimental setup consists of a $^{87}\mathrm{Rb}$  cold atomic ensemble, a polarization-independent quantum frequency converter in Sagnac configuration, spooled fibers and a polarization analysis module with superconducting nanowire single photon detectors (SNSPDs). The detailed experimental setup is shown in Fig. \ref{Fig1}.

\paragraph{Atom-photon generation}

The atoms in the ensemble are laser cooled through a three-dimensional magneto-optical trap, which gives a temperature of  ${\sim}$10 $\mu$K after polarization gradient cooling. In each cycle, atoms are initialized to $\left |  g \right \rangle = \left | 5S_{1/2}, F=1, m_{F}=-1  \right \rangle $ via optical pumping with an efficiency $\sim$90\%. A 20-MHz blue-detuned short write pulse ($\sim$50 ns) couples state $\left |  g \right \rangle$ and state $\left |  e \right \rangle = \left | 5P_{3/2}, F=2, m_{F}=0  \right \rangle $, starting the excitation process as shown in the inset of Fig. \ref{Fig1}, {see more details in {Supplemental Material \cite{SupplementalMaterial}} part 1}. In subsequent {spontaneous Raman scattering (SRS)}, atomic spin wave and the polarization state of the scattered photon (hereafter the photon is defined as write-out photon) establish the maximally entangled atom-photon state as follows:

\begin{equation}
\begin{split}
\ket{\Phi}_\mathrm{atom-photon} = \frac{1}{\sqrt{2} } \left ( \left | \Downarrow    \right \rangle _z \left | L  \right \rangle - \left | \Uparrow    \right \rangle _z \left | R  \right \rangle \right ) \\
\label{Eq.1.1}
\end{split}
\end{equation}

where $\left | L \right \rangle $ and $\left | R \right \rangle $ donate left-circular ($\sigma^+$) and right-circular ($\sigma^-$) photonic polarization states, respectively, {and $-$ symbol comes from the sign of the Clebsch-
Gordan (CG) coefficients.} $\left | \Downarrow \right \rangle _z = \frac{1}{\sqrt{N}} \sum_{j}^{N}e^{i{\bm{k}}\cdot{\bm{r}}_{j}}|g...{\downarrow_{j}}...g\rangle$ and $\left | \Uparrow  \right \rangle _z  = \frac{1}{\sqrt{N}} \sum_{j}^{N}e^{i{\bm{k}}\cdot{\bm{r}}_{j}}|g...{\uparrow_{j}}...g\rangle$ show the atomic spin wave in which $N$ is the number of atoms, with $\left |  \downarrow \right \rangle = \left | 5S_{1/2}, F=2, m_{F}=-1  \right \rangle $ and $\left |  \uparrow \right \rangle = \left | 5S_{1/2}, F=2, m_{F}=+1  \right \rangle $ are atomic states along the quantization axis direction ($z$ direction), $\bm{k}=\bm{k}_\mathrm{w}-\bm{k}_\mathrm{wo}$ corresponds to the  wave vector of the atomic spin wave and equals to the difference between wave vector of write beam ($\bm{k}_\mathrm{w}$) and wave vector of write-out photons ($\bm{k}_\mathrm{wo}$), $\bm{r}$ is the position of the excited atom.

\begin{figure}
  \centering

  \begin{minipage}[b]{.48\columnwidth}
    \centering
    \includegraphics[width=\linewidth]{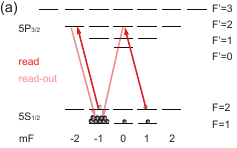}
    \vspace{-2pt}
  \end{minipage}\hfill
  \begin{minipage}[b]{.48\columnwidth}
    \centering
    \includegraphics[width=\linewidth]{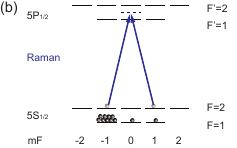}
    \vspace{-2pt}
  \end{minipage}

  \vspace{4pt}

  \begin{minipage}[b]{.48\columnwidth}
    \centering
    \includegraphics[width=\linewidth]{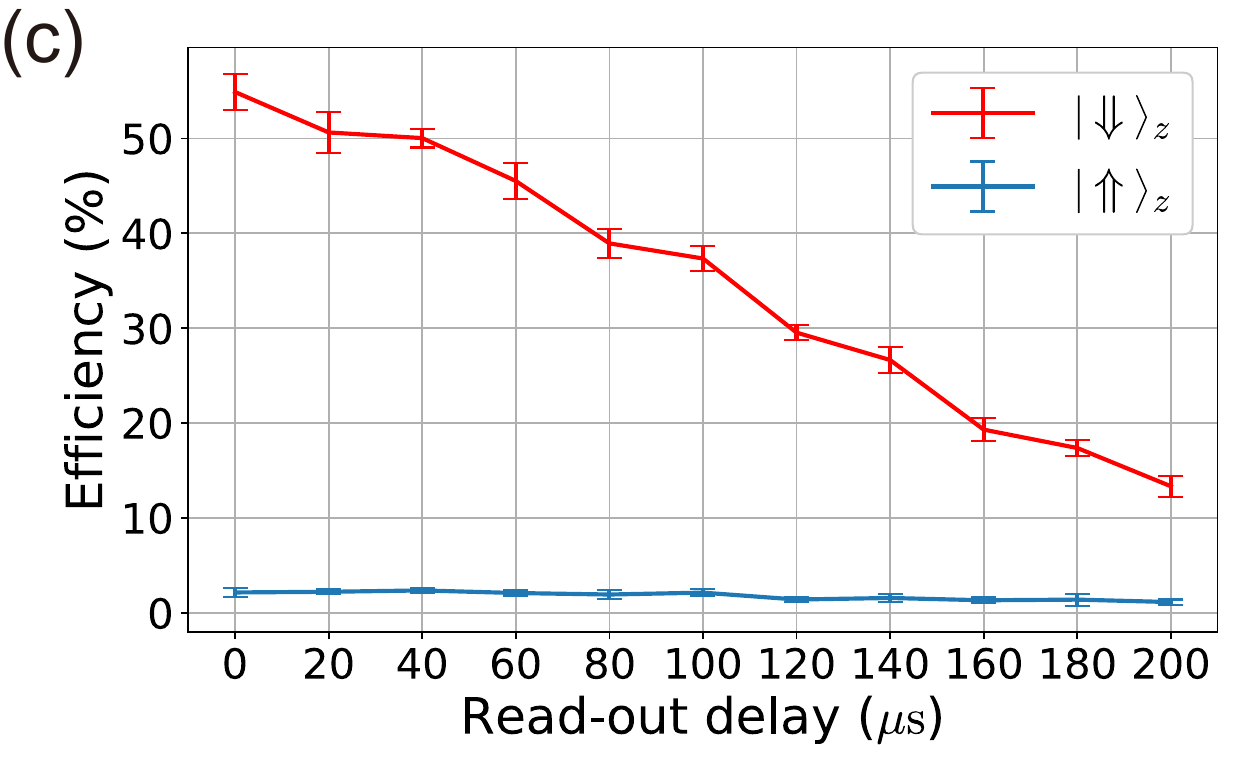}
    \vspace{-2pt}
  \end{minipage}\hfill
  \begin{minipage}[b]{.48\columnwidth}
    \centering
    \includegraphics[width=\linewidth]{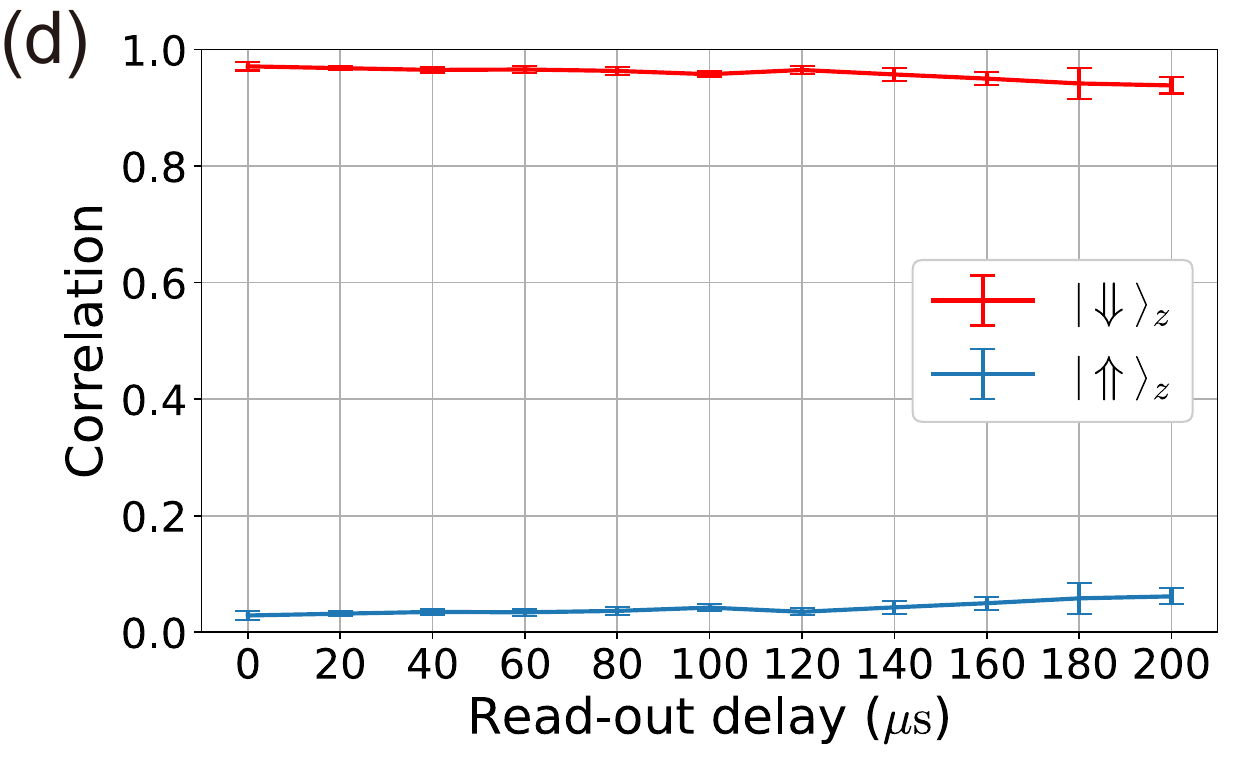}
    \vspace{-2pt}
  \end{minipage}
  \caption{Read scheme and performance. (a) For read in atomic $\sigma_z$ basis, the $\sigma^-$ polarized read pulse, with a duration of approximately 250 ns, couples states $\left | 5S_{1/2}, F=2, m_{F}  \right \rangle$ and $\left | 5P_{3/2}, F=2, m_{F}-1  \right \rangle$ to convert the atomic spin wave into a single photon for detection. 
  (b) For read in superposition basis, a $\pi /2$ Raman pulse with a duration $\sim$400 ns containing simultaneous $\sigma^+$ and $\sigma^-$ polarization components is applied before the read pulse. The Raman pulse couples the $\left | 5S_{1/2}, F=2 \right \rangle$ $\to $ $\left | 5P_{1/2}, F=1 \right \rangle $  and $\left | 5P_{1/2}, F=2 \right \rangle $ transitions with an intermediate detuning of 407 MHz to map the states in $\sigma_x$ basis to $\sigma_z$ basis, namely $\left | \Downarrow  \right \rangle_x = \left | \Downarrow  \right \rangle_z + \left | \Uparrow  \right \rangle_z $ $\to $ $\left | \Downarrow  \right \rangle_z$ and  $\left | \Uparrow  \right \rangle_x = \left | \Downarrow  \right \rangle_z - \left | \Uparrow  \right \rangle_z $ $\to $ $\left | \Uparrow  \right \rangle_z$, achieving measurement on the superposition basis of atoms. (c-d) The change of internal retrieval efficiency and correlation with memory time when read-out photon fixed to $\sigma^-$ and write-out projected on $\sigma^-/\sigma^+$. {Error bars are measured one standard deviations.}
  }
\label{Fig2}
\end{figure}

\paragraph{Atomic state readout}
To verify atom-photon entanglement, the write-out photons are projected and detected by APDs (efficiency 65\%, {noise counts $\sim$200 Hz,} {two filtering cavities with a total ~75\% efficiency are implemented}) for local measurement (or by SNSPDS for long-fiber measurement). If the detection is successful then a read pulse with $\sigma^-$ polarization is applied to convert the atomic spin wave to read-out photons for detection, as depicted in Fig. \ref{Fig2}(a). The $\left | \Uparrow  \right \rangle _z$ state is read out as $\sigma^+$-polarized photon, while the $\left | \Downarrow  \right \rangle _z$ state is read out as $\sigma^-$-polarized photon. Due to difference in CG coefficients, the read-out efficiencies for $\left | \Downarrow  \right \rangle _z$ and $\left | \Uparrow  \right \rangle _z$ are imbalanced, preventing direct read out in atomic superposition basis. To overcome this limitation, Raman state transfer is here implemented using a $\pi$/2 pulse to map the atomic state in superposition basis $\sigma_x$ into the eigenbasis $\sigma_z$ in Fig. \ref{Fig2}(b), together with a guiding field of 127.5 mG along $z$ direction, we can project any atomic superposition within a Larmor period ($\sim$5.6 $\mu$s),  {see more details in {Supplemental Material \cite{SupplementalMaterial}} part 2}.

{Former results of atom-photon entanglement in the cavity scheme with a longest lifetime of 458 ms reports an internal retrieval efficiency of 58\% \cite{Bao-2021-PRL}, while highest internal retrieval efficiency of 88\% is achieved with only 75-$\mu$s memory lifetime \cite{JingBo-2019-NaturePhotonics}.}
In our cavity-free scheme, the OD of the atomic ensemble is optimized to 10. 
Under this OD and phase-matching conditions, the readout efficiency is substantially enhanced due to the collective effects. 
{As shown in Fig. \ref{Fig2}(c), the atomic memory reaches a 1/e lifetime of 160 $\mu$s and  an internal retrieval efficiency of $\sim$55\%, significantly higher than the 35\% achieved in prior study \cite{JingBo-2019-NaturePhotonics}.}
{The end efficiency on APD is around 19\% including mode-matching efficiency 73\% (including mirror losses, fiber coupling loss and fiber-fiber connection loss), two filtering-cavity efficiency 73\% and APD efficiency 65\%.} 

The lifetime here is mainly limited by motion induced decoherence as each atomic spin wave will contribute to different phase item $e^{i{\bm{k}}\cdot{\bm{r}}_{j}}$.
When measured locally, the contrast in $\left | \Downarrow  \right \rangle _z$/$\left | \Uparrow  \right \rangle _z$  basis is kept higher than 90\% within memory lifetime and the initial fidelity of the atom-photon entanglement is estimated to be around 96\%.

\section{Results}

\paragraph{Long-distance atom-photon entanglement}
Based on all the aforementioned setups and methods {and with the aid of a high-efficiency polarization-independent (PI) QFC module (see End matter PIQFC module)}, we can now distribute atom–photon entanglement over telecom fibers and record interference fringes at lengths $L$ of 10 m, 5 km, 10 km and 20 km, a local measurement with $L$ = 10 m for 780-nm photon transmission is also provided ({end excitation probability 0.9\% }). Notice that a single-trip readout delay is implemented here: the atomic state is read out immediately upon photon-detection events. The related results are shown in Table \ref{table1}.

\begin{table*}[t]
\caption{\label{table1} Verification of atom-photon entanglement under different configurations. The measurement differs mainly in fiber length, photon wavelength, atomic readout delay (referenced to start of the write electric pulse) and excitation {probability}. Visibility in both the eigen basis and superposition basis are provided.}
\begin{ruledtabular}
\begin{tabular}{cccccc}
 Fiber length $L$  &10 m & 10 m &  5 km &  10 km &	 20 km\\ 
\hline
 Visibility in $\ket{\Uparrow_z}$/$\ket{\Downarrow_z}$ (\%)&95.5±0.85&	92.8 ± 1.6 &	95.1 ± 2.6 &	93.8 ± 3.0 &	89.0 ± 5.7 \\
 Visibility in $\ket{\Uparrow_x}$/$\ket{\Downarrow_x}$ (\%) &94.2±0.46&	94.1 ± 2.7 &	91.6 ± 2.7 &	88.9 ± 4.1 &	83.6 ± 9.3 \\
 Photon wavelength  & 780 nm& 1522 nm& 1522 nm& 1522 nm& 1522nm\\
 Readout delay  &0.7 $\mu$s& 1.10 $\mu$s &  25.65 $\mu$s &  50.15 $\mu$s &	 99.25 $\mu$s\\
SNR (exc probability (\%)) & 150 ({1.8})& 44 ({1.5}) & 59 ({1.9})& 56 (({1.8})& 89 ({3.0})\\
 Repetition rate  &31.4 kHz& 31.0 kHz &  5.8 kHz &  3.4 kHz &	 1.7 kHz \\

{End read-out efficiency (\%)} &{19.0±1.2}   &{18.5±1.1}  &{17.8±1.7}  &{18.0±1.5} &{15.8±2.3}  \\
\end{tabular}
\end{ruledtabular}
\end{table*}

Usually, to have a low two-photon events, the excitation {probability, defined as the probability of a photon scattered into the fiber with one excitation try} is set to very low (only around 1\%-2\%).
As is shown in the Table \ref{table1}, only for $L$ = 20 km, the excitation {probability} is set a bit higher to get more events despite with slightly reduced visibilities.
In general, the measurement in superposition basis shows a lower visibility and faster decay compared to that in the eigen basis, which is attributed to two factors. The first is the imperfect Raman state transfer, even the power of the Raman beam is locked (drift far below 1\%), the guiding field introduces a small two-photon detuning for the Raman transfer. This results in a $\sim$1.5\% drop in  visibility when measured locally. 
The second contribution arises from imperfect magnetic-field control: because the interference is read out via Larmor precession, any field instability directly degrades the visibility.

\begin{figure}
\centering

  \begin{minipage}[b]{.48\columnwidth}
    \centering
    \includegraphics[width=\linewidth]{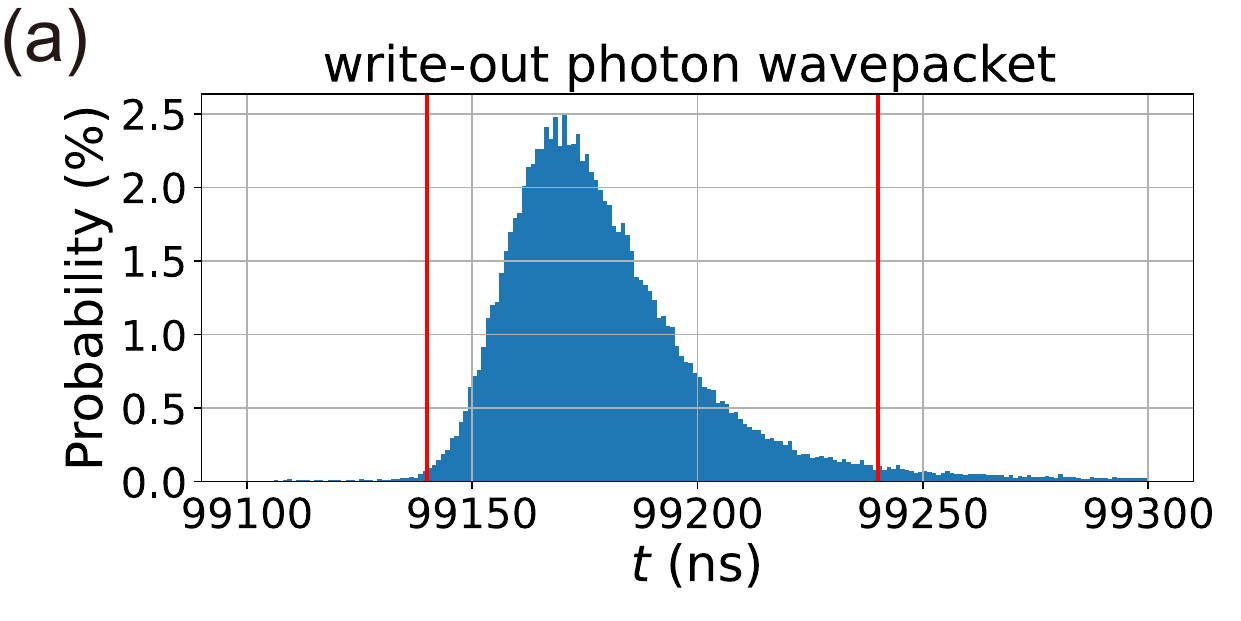}
    \vspace{-2pt}
  \end{minipage}\hfill
  \begin{minipage}[b]{.48\columnwidth}
    \centering
    \includegraphics[width=\linewidth]{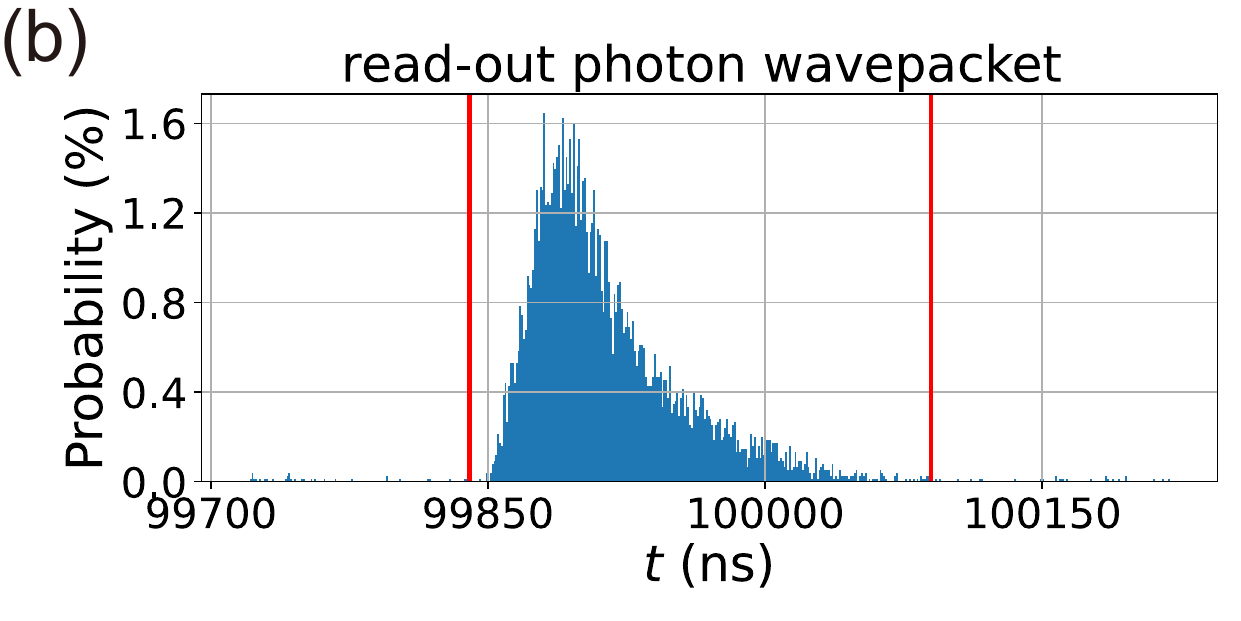}
    \vspace{-2pt}
  \end{minipage}

  \vspace{2pt}
  
  \begin{minipage}[b]{.48\columnwidth}
    \centering
    \includegraphics[width=\linewidth]{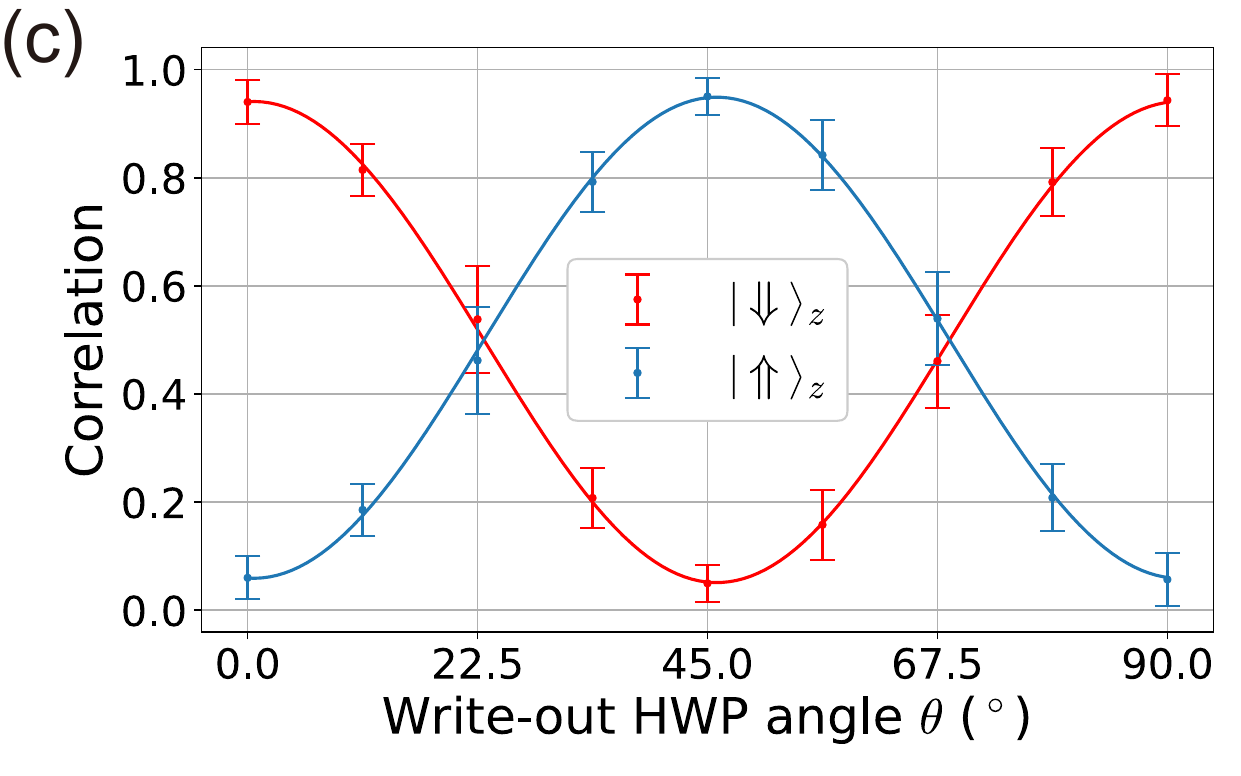}
    \vspace{-2pt}
  \end{minipage}\hfill
  \begin{minipage}[b]{.48\columnwidth}
    \centering
    \includegraphics[width=\linewidth]{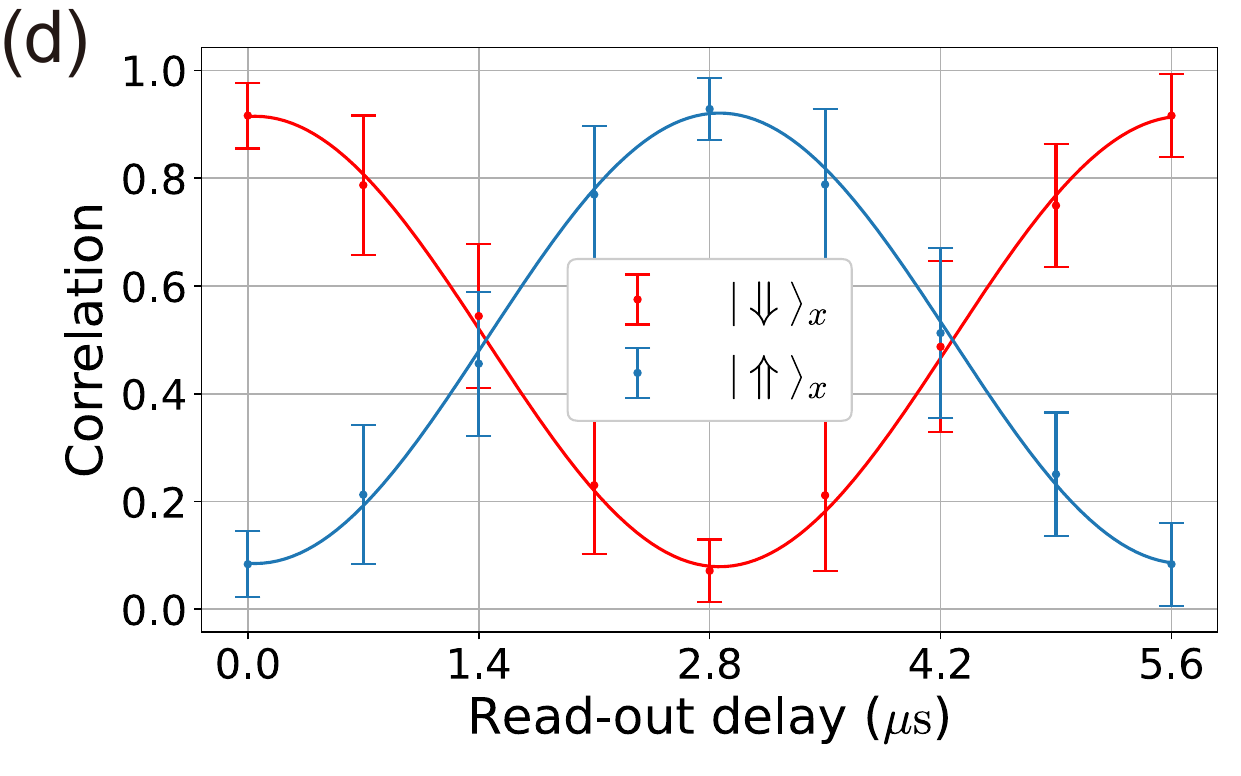}
    \vspace{-2pt}
  \end{minipage}

\caption{{Atom-photon performance with 20-km fiber transmission.} (a-b) The time histogram of 1552-nm write-out photons and 780-nm read-out photons. In both figures, red vertical lines demarcate the full width signal window. Each bin of the histogram occupies 1 ns. (c-d) The normalized coincidence counts for changing the angle of a HWP under eigen basis and changing the read-out delay (offset $\sim$100$\mu$s) under superposition basis. The results are fitted by a sinusoidal function. {Error bars are measured one standard deviations.}
}
\label{Fig3}
\end{figure}

\paragraph{20 km atom-photon entanglement}
Fig. \ref{Fig3}a and \ref{Fig3}b display the wave packet of write-out photons after 20 km  optical fiber transmission, and  wave packet of read-out photons. The write-out and read-out signals exhibit signal to noise rations (SNRs) of 89 and 597, respectively. 
Fig. \ref{Fig3}c is the interference in $\ket{\Uparrow_z}$/$\ket{\Downarrow_z}$ basis, measured  over a period of 15 hours, in total 78288 successful excitation events and 7353 coincidence events were observed, with an average excitation probability of 0.205\% on SNSPDs, {see more details in {Supplemental Material \cite{SupplementalMaterial}} part 4}.   Fig. \ref{Fig3}d is the interference in $\ket{\Uparrow_x}$/$\ket{\Downarrow_x}$ basis, calculated over 80286
successful excitation events and 6043 coincidence events. 
These counts are primarily limited by the low repetition rate due to the 20-km long-distance transmission, along with the loss from QFC and fiber transmission and time consumed for fiber polarization compensation and filtering-cavity temperature auto optimization.
The corresponding visibilities can be found in Table \ref{table1} with $L$ = 20 km, {with the estimated quantum bit error rate of 5.54±4.1\% in eigen basis and 7.96±6.56\% in superposition basis}. The error bar of corelation in superposition basis as shown in Fig. \ref{Fig3}d is larger due to the imperfect control of the magnetic field. This becomes particularly severe for over-night measurement because the magnetic-field sensor drifts over time and temperature (3\%).

\begin{figure}
\centering
\includegraphics[width=8cm]{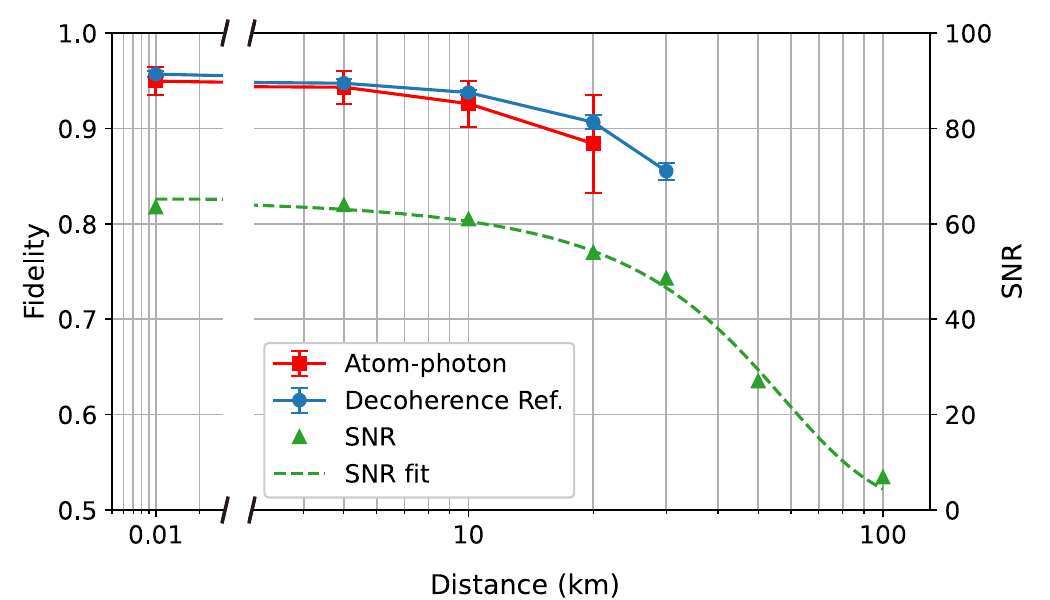}
\caption{Fidelity and SNR dependence on transmission distance. Red squares denote atom-photon entanglement fidelity at different transmission distances for the telecom measurement, while blue circles represent fidelity for local atom-photon entanglement but with same readout delay (the red and blue lines are plot for visual reference). The green triangles  are measured SNR with same exc efficiency (1\% on local APDs), the fitted curve is from the SNR model. {Error bars are measured one standard deviations.}
}
\label{Fig4}
\end{figure}

\paragraph{Fidelity}
To quantitatively describe atom-photon entanglement, we estimate the fidelity $\mathcal{F} = \frac{1}{4} (1+V_{\ket{\Uparrow_z}/\ket{\Downarrow_z}}+2V_{\ket{\Uparrow_x}/\ket{\Downarrow_x}})$. Fig. \ref{Fig4} presents the dependence of fidelity on transmission distance (red squares), a local measurement with the same readout delay is labeled as decoherence reference (blue circles) {and a separate measurement of SNR dependence on fiber distance and the fit are also depicted (see End matter SNR section).}

{Currently, the initial fidelity is around 96\%, in which the visibilities both in eigen basis or superposition basis are around 94-95\%, this infidelity comes from mainly two-photon excitations, imperfect state preparation, slight polarization errors and noise counts on APDs (around 200 CPS, including environmental photons and dark counts). For long fiber-measurements with QFC, we can see  the SNR of converted 1522-nm photon is quite high and the fidelity is very close to local reference, which means we are not limited by SNR so far but limited by atomic decoherence.}

Owing to the high repetition rate of local measurements (31 kHz, {defined as the number of excitation tries per second}), data for each point are collected within just one to two hours—a window short enough that magnetic-field sensor drift is negligible, yielding exceptionally small error bars. With PIQFC, we observe an entanglement fidelity between the atoms and telecom photon exceeding  80\% after photon transmission over 20-km fiber, the remaining infidelity being dominated by atomic decoherence.

\section{Outlook and conclusion}
{The next milestone towards quantum network is to establish distant entanglement between two cavity-free atomic memories, where entanglement swapping can be performed in the middle station to herald the entanglement \cite{Zhang-2022-Nature}. 
Based on the result of current setup, we estimate the performance of a potential 20-km atom-atom entanglement with a 10-km atom-photon distribution distance each side and the implementation of round-trip readout delay for both sides.}

{The atom-atom entanglement probability is estimated as 
$(0.2\%)^{2}\times1.575^2/2=4.96\times10^{-6}$ where 1.575 corresponds rate increase due to photon transmission of 10 km instead of 20 km with a loss of 0.2 dB/km, and the division by two represents the efficiency of BSM. With the repetition rate of 1.7 kHz, then the entanglement generation time would be around 119 seconds. The estimated averaged visibility of atom-atom entanglement would reach $V=((0.89+0.836)/2)^2=0.745$, resulting in a fidelity around 0.8. Based on that, the link efficiency \cite{Humphreys-2018-Nature}, which is defined as the ratio of entanglement-generation rate to the decoherence rate, is calculated as $1.35\times10^{-6}$ excluding the end readout efficiencies of $\sim$16\% for both sides.}

{To improve the link efficiency, the first idea is to increase the memory lifetime. With an optical lattice, the lifetime can be extended to 458 ms \cite{Bao-2021-PRL}, however with the repetition rate drop to around $1/3$ to $1/2$ due to longer time for preparing ensembles with lattice. In addition, the excitation probability can be increased by a factor of 2, albeit with a slight fidelity loss. Together with the filtering-cavity transmission improvement, this yields an overall enhancement by a factor of 5 to 10. In total, an improvement factor of $4.8\times10^3$ is highly achievable. A further extension of coherence time to a minute level is possible with extremely high vacuum and dynamic decoupling protocol \cite{Kuzmich-2013-PRA}, and ultra-low-loss telecom fiber (0.17dB/km) can also be used to enhance the efficiency \cite{Luo2025Entangling}.}

{
On the atomic detection side, the end detecting efficiency for readout could be improved further by increasing the OD of the atomic ensemble, which can be achieved by a spatial dark-line technique\cite{Ketterle-1993-PRL} or magnetic-field compression \cite{Gao-2025-AdvancedPhotonics}. More efforts on filtering module or applying high-efficiency SNSPDs ( $\ge$90\% efficiency) instead of APDs will also enhance the end readout efficiency.}

In this work, by utilizing PIQFC with 48.5\% external efficiency and atomic-ensemble memory with approximately {55}\% initial retrieval efficiency and 160 $\mu$s lifetime, atom-photon entanglement with fidelity $\mathcal{F} > 0.80$ is observed over 20 km of optical fiber. An upgrade of the system will immediately enable the atom-photon distribution of 100 km, however for heralded long-distance atom-atom Bell entanglement, a longer memory lifetime and a higher memory efficiency are required, { otherwise the atomic state has to be readout immediately after the excitation\cite{Bao-2020-Nature}}.

\section{Acknowledgements} We thank Dr. Bo Jing  from SWJTU and Dr. Jun Li from USTC for very helpful discussion regarding the atom-photon generation module. 
We acknowledge funding by the Quantum Science and Technology-National Science and Technology Major Project (Grant No. 2021ZD0301102), National Natural Science Foundation of China (Grant No. 12104361).

\begin{acknowledgments}

\end{acknowledgments}

\bibliography{apscication}

\appendix
\section*{End Matter}
\paragraph{PIQFC module}

\begin{figure}[ht!]
\includegraphics[width=8cm]{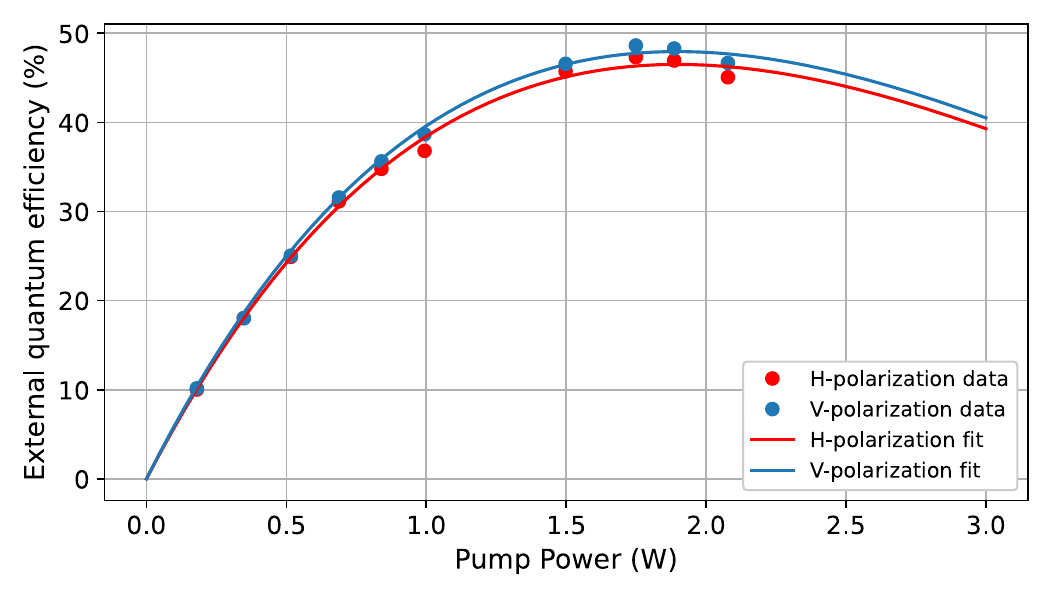}
\caption{ The QFC module's external quantum efficiency versus pump power. The data points are fitted with 
$\eta(P)=\eta_{\max} \sin^2\left(\sqrt{\alpha_{\text{nor}} P} L\right)$\cite{Wong-2004-Opt.Lett.}.
The pump power is set to 1.749 W, with maximized efficiency of 47.2\% and 48.5\% for the H- and V-polarization arms, respectively.
}
\label{Fig5}
\end{figure}

To distribute the atom-photon entanglement over a distance, a polarization-independent quantum frequency conversion (PIQFC) module is implemented here\cite{Ikuta-2018-Nat.Commun., Chen-2024-Phys.Rev.Appl.}. 
The system employs a pump-enhanced cavity scheme within the Sagnac interferometer, achieving high conversion efficiency through difference-frequency generation in a PPLN bulk crystal, and a filtering module is designed to reduce noise primarily coming from the QFC converter. The external quantum efficiency (EQE) of both arms with different pump laser powers is shown in Fig. \ref{Fig5}. The EQE of the frequency conversion system, including the converter and filtering module, is defined from the 780-nm photon input port to the reflection of the filter module's VBG (the polarization analysis module and last coupling before SNSPs are excluded), and the max EQE equals 48.5\%. {See more details in {Supplemental Material \cite{SupplementalMaterial}} part 3}.

With the aid of the high-efficiency PIQFC module, the converted 1522-nm photons can propagate through long fibers up to 20 km, where the polarization drift of the long fibers is compensated once per several minutes automatically. The following step is polarization projection, including detection by SNSPDs with an efficiency $\sim$88\% and dark counts $\sim$30 counts per second (CPS). The detection events, which trigger the atomic state readout, are accepted only within a 200-ns hardwired timing window, allowing correlation measurements for the entanglement.

\paragraph{SNR}
The SNR of the telecom photons is primarily limited by the QFC module and dark counts of the SNSPDs. Here, the noise counts from QFC start at 270 CPS/channel, and dark counts of detectors are 30 CPS/channel constantly. Fortunately, the noise photons generated from the QFC undergo the same attenuation as the signal photons in fibers, so their relative level remains unchanged, leaving the dark counts the only issue for SNR with long-fiber transmission. 

We further set up a separate measurement to see the dependence of SNR of write-out photons with transmission distance when setting excitation {probability} constant 2\% (1\% on local APDs). The corresponding SNR calculated with full width is list in Fig. \ref{Fig4} as green triangles. We can see that the low-noise and high-efficiency QFC gives a SNR of 6.9 for transmitted photons with fiber length
up to 100 km. If we take FWHM of the wave packet for calculation, although total event rate will drop to approximately 70\%, the SNR will be even doubled.
In addition, a model for SNR simulation over fiber length is shown in Fig. \ref{Fig4} as the green dashed line, 
{see more details in {Supplemental Material \cite{SupplementalMaterial}} part 7}.

\end{document}


\preprint{APS/123-QED}

\title{Supplementary Material for Long-distance distribution of atom-photon entanglement based on a cavity-free cold atomic ensemble }
\author{Tian-Yu Wang$^{1,2,3,\ddag}$, Ren-Hui Chen$^{1,2,3,\ddag}$, Yan Li$^{1,2,\ddag}$, Ze-Hao Shen$^{4,\ddag}$, Xiao-Song Fan$^{1,2,3}$, Zheng-Bang Ju$^{1,2,3}$, Tian-Ci Tang$^{1,2,3}$, Xia-Wei Li$^{1,2,3}$, Jing-Yuan Peng$^{1,2,3}$, Zhi-Yuan Zhou$^{1,2,3,*}$, Wei Zhang$^{3,\dag}$, Guang-Can Guo$^{1,2,3}$, Bao-Sen Shi$^{1,2,3,\S}$}
\affiliation{
$^1$Laboratory of Quantum Information, University of Science and Technology of China, Hefei 230026, China\\
$^2$Anhui Province Key Laboratory of Quantum Network,
University of Science and Technology of China, Hefei 230026, China\\
$^3$Hefei National Laboratory, University of Science and Technology of China, Hefei 230088, China\\
$^4$School of Physics, Xi’an Jiaotong University, Xi’an 710049, China
}




\date{\today}

\maketitle

\renewcommand{\thefigure}{S\arabic{figure}} 

\section{1. Platform and sequence}
The vacuum system comprises an ion getter pump maintaining high vacuum, Rb dispensers for atom generation, and a glass vacuum chamber with a size of 100 mm$\times$35 mm$\times$35 mm. In the experiment, the dispenser is driven at 3.2 A to supply sufficient $^{87}$Rb atoms while maintaining vacuum pressure below $2\times 10^{-8}$Pa based on discharge current measurement.

The experimental sequence begins with a 28.6-ms cooling for atom capture, employing three 14-mm diameter beams (retro-reflection structure) of 12 mW each for cooling (20-MHz red detuned to transition $ \left | 5S_{1/2},  F=2\right \rangle$ $\to$ $\left | 5P_{3/2}, F=3\right \rangle $), a 34-mW beam for repump process (on resonant with transition $ \left | 5S_{1/2},  F=1\right \rangle$ $\to$ $\left | 5P_{1/2}, F=2\right \rangle $), and a gradient magnetic field $\sim$12 G/cm. During the final 3 ms of this phase, the gradient magnetic field is switched off (120-$\mu$s 1/e time ) while the detuning of cooling light is increased and the power reduced (polarization gradient cooling, PGC), resulting in an atomic ensemble with a temperature of $\sim$10 $\mu$K. At the end of cooling, we apply a 16-$\mu$s long optical-pumping process to initialize all atoms on  $\left | 5S_{1/2}, F=1 \right \rangle$ with an efficiency 100\% or on $\left | 5S_{1/2}, F=1, mF=-1 \right \rangle$ with an efficiency $\sim$90\%.

Subsequently, an experimental phase of up to 6 ms commences, consisting of 4 identical rounds (in between we are planning to apply a PGC process, but in this experiment, it is only 2-$\mu$s delay) with many excitation trials in each. In the 6-ms duration, the OD with an initial value of 10, measured by left-circular or right-circular probe transition $\left | 5S_{1/2}, F=1 \right \rangle$ $\to $ $\left | 5P_{3/2}, F'=1 \right \rangle $ along write-out path, is decreased by less than 1/3, mainly due to atomic diffusion and gravitational fall. 

In each round, there are up to 250 cycles (the number varies with transmission distance, e.g. 15 cycles for 20-km measurement), where each cycle consists of a short 2-$\mu$s optical pumping process for state preparation, a 50-ns short write pulse and a delayed 200-ns hardwired photon window to accept single-photon pulses from detectors at correct timing. If any detector click happens in the photon window, a read process with a 250-ns read laser pulse will be triggered. Only after all 4 rounds (1000 cycles for 0-km measurement, 60 cycles for 20-km measurement) have been completed do we recapture the atoms and repeat the above processes.

\section{2. Magnetic field control}
Besides MOT coils, there are three pairs of compensation coils employed here to set a uniform magnetic field for the atomic ensemble. Each pair of square coils with side length around 20 cm forms cuboid geometry rather than cubic geometry to get a more uniform field. A magnetic-field sensor, which is triggered by the ending of cooling and takes 1 ms to sample the field in three directions, is placed as close as possible to the vacuum chamber to detect the residual field near atoms. Together with a digital-to-analog converter and a home-made voltage-controlled current source, a feedback servo for the magnetic field is achieved with a field deviation below 1 mG along three directions. 

Although the feedback is slow, updated only once per cooling cycle ($\sim$33 ms), it is still sufficient to suppress both the geomagnetic field ($\sim$600 mG) and disturbances from the subway (Vpp$\sim$10 mG along Y direction). The only problem is the residual mutual field from inductance, which is around 6 mG amplitude on sensor along Y direction caused by fast switching off of the MOT coils. In this experiment, we apply a constant 127.5-mG guiding Z field to suppress the induced time-varying magnetic field, but still one can see small oscillations on readout efficiency decay as shown in Fig. 2(c) in the main text. A higher guiding field is possible but will make cooling process less efficient, resulting in a higher atomic temperature.

\section{3. PIQFC converter}

\begin{figure}[ht!]
\centering
\includegraphics[width=8cm]{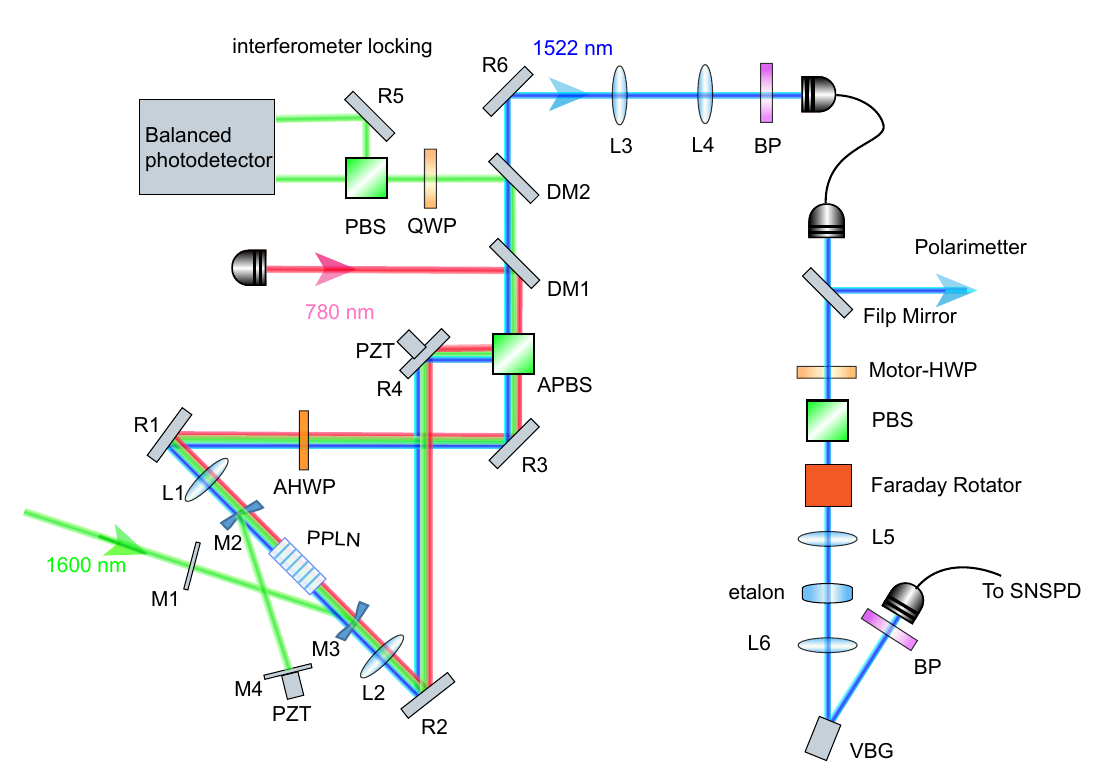}
\caption{ PIQFC module setup. The pump enhancement cavity consists of four mirrors (M1, M2, M3, and M4). M4 is fixed on a piezoelectric transducer (PZT) to control the cavity length. Lenses (L1, L2) are with anti-reflective coating at 1522-nm and 780-nm. R1, R2, R3, and R4 are highly reflective coating reflectors at 780-nm and 1522-nm. R4 is fixed on a PZT to control the interferometer length. AHWP, achromatic half-waveplate. APBS, achromatic polarizing beam splitter. DM1, dichroic mirror for 780-nm reflection and 1522-nm,1601-nm transmission. DM2, dichroic mirror for 1601-nm reflection and 1522-nm transmission. BP, 10-nm bandpass filter (BPF) centered at 1520 nm. PBS, polarizing beam splitter.  QWP, quarter-waveplate. Motor-HWP, motor half-waveplate. R5 R6, reflectors at C band. VBG, volume Bragg grating. SNSPD, Superconducting Nanowire Single Photon Detector.
}
\label{FigS1}
\end{figure}

Recently, several polarization-independent quantum frequency conversion configurations have been developed to eliminate the polarization dependence of the second-order nonlinear process. The Sagnac interferometer configuration has been widely investigated because of its benefits, such as compact component design and passive phase stabilization. We use a type-0 quasi-phase-matched periodically poled lithium niobate (PPLN) bulk crystal within a Sagnac interferometer to achieve difference-frequency generation (DFG), converting 780-nm photons to 1522-nm photons at a temperature of 60°C.
The experimental setup is shown in Fig. \ref{FigS1}. The PPLN crystal has a length of 50 mm and a poling period of 19.65 $\mu$m. Although bulk crystals typically require higher pumping power than waveguide crystals to achieve near-unity internal conversion efficiency, they provide notable advantages such as eliminating coupling losses and reducing noise levels.

We demonstrate a 94.5\% internal quantum efficiency in a bulk crystal using cavity-enhanced DFG process, which requires lower power compared to direct pumping. A figure-eight-shaped 1601-nm pump-enhancement standing-wave cavity with a finesse of about 160 is in the interferometer. The cavity is locked using the Pound-Drever-Hall (PDH) technique. The pump light is generated by a diode laser and amplified with an erbium-doped fiber amplifier, which is stabilized at 1601.1451-nm wavelength using a wavelength meter. 

The Sagnac interferometer consists of an APBS, an AHWP, four mirrors, and a PPLN crystal. The 780-nm signal photons are reflected by DM1 into the Sagnac interferometer. The 780-nm input signal light is split into two arms using an achromatic PBS. The V-polarized signal light and the 1601-nm pump light are mixed in the PPLN crystal to generate V-polarized 1522-nm light, then the converted light is rotated to H polarization via an achromatic HWP. For the H-polarized input signal photon, it is first rotated to V polarization by the same HWP and then converted to V-polarized light at 1522 nm. Finally, the two converted light beams coherently superimpose at the PBS for output. The converted light passes through the DM2 and is coupled into the fiber via the lens assembly. Since the pump light does not co-propagate with the signal light through the interferometer loop, the Sagnac interferometer loses its intrinsic phase stability, necessitating active path-length stabilization.
We set $L_1$ to represent the path length from the APBS transmission port (input 780-nm signal light direction) to the crystal center, and $L_2$ to represent the path length from the APBS reflection port to the crystal center. $S_1$ represents the path length from the cavity mirror M1 to the center of the crystal in the counterclockwise direction, and $S_2$ represents the path length from the cavity mirror M4 to the center of the crystal in the clockwise direction. Considering that the phase of the converted light from the DFG process is determined by the pahse of pump light and the signal light, the phase $\varphi_{conH}$ of the H-polarized  converted light at the interferometer output is given by:
\begin{eqnarray}
    \varphi_{conH}=k_{s}L_2-k_{p}S_1+k_{c}L_1+\varphi_{dispH}
\end{eqnarray}
And for the phase $\varphi_{conV}$ of V-polarized conversion light:
\begin{eqnarray}
    \varphi_{conV}=k_{s}L_1-k_{p}(S_1+2S_2)+k_{c}L_2+\varphi_{dispV}
\end{eqnarray}
${{k}_{s}}$ is the wavevector of the 780-nm signal light, ${{k}_{c}}$ is the wavevector of the 1522-nm conversion light, and ${{k}_{p}}$ is the wavevector of the 1601-nm pump light. ${{\varphi }_{dispH/V}}$ is the extra phase attributed to the H/V polarized light passing through the  dispersive optical components. To calculate the phase difference $\Delta\varphi=\varphi_{conH}-\varphi_{conV}$ considering ${{k}_{p}}={{k}_{s}}-{{k}_{c}}$ in free space, we get:
\begin{eqnarray}
    \Delta\varphi={{k}_{s}}(2S_2+L_2-L_1)+\Delta{\varphi_{disp}}
\end{eqnarray}
$\Delta{\varphi_{disp}}=\varphi_{dispH}-\varphi_{dispH}$ is the phase affected by wavelength and temperature. Considering light wavelength locking and laboratory temperature control technology, it can be considered a constant. We find that the phase difference depends primarily on the interferometer and cavity arm lengths, so we use the weak pump light leaked through M2 and M3 to actively stabilize the path length. We also calculate the phase $\varphi_{M2}$ of the pump light transmitted by M2 at the interferometer output: $\varphi_{M2}={{k}_{p}}(S_1+L_1)+\varphi_{dispM2}$. And for the pump light transmitted by M3, we have $\varphi_{M3}={{k}_{p}}(S_1+2S_2+L_2)+\varphi_{dispM3}$. So the phase difference $\Delta\varphi_{p}=\varphi_{M3}-\varphi_{M2}$ is:
\begin{eqnarray}
    \Delta\varphi_{p}={{k}_{p}}(2S_2+L_2-L_1)+\Delta\varphi_{disp-pump}
\end{eqnarray}
The $\varphi_{dispM2}$ and $\varphi_{dispM3}$ represent the extra phase attributed to the leaked pump light passing through the dispersive optical components, and $\Delta\varphi_{disp-pump}=\varphi_{dispM3}-\varphi_{dispM2}$. Similarly, the $\Delta\varphi_{disp-pump}$ can be considered a constant.
Therefore, we can actively stabilize the path length using leaked pump light. The pump light transmitted by M2 and M3, and reflected at DM2 to enter the interferometer locking module, including a QWP, a PBS, and a balanced photodetector, as shown in Fig. S1. The output power of the PBS transmitted port $P_{H}$ is $P_{pump}(1+sin\Delta\varphi_{p})$,$P_{pump}$ is the power of pump light transmitted by M2 and M2, and for the reflected port:$P_{V}$ is $P_{pump}(1-sin\Delta\varphi_{p})$. Now the balanced photodetector outputs the differential error signal $P_{error}$:
\begin{eqnarray}
    P_{error}=P_{H}-P_{V}=2P_{pump}sin\Delta\varphi_{p}
\end{eqnarray}
It serves as an error signal for the PID lock circuit. The feedback on the path length is applied on the PZT attached to R4.

The converted 1522-nm photons passing through DM1, are separated from the weak 1601-nm interferometer locking light by DM2, and then coupled into a single-mode fiber. A BPF is placed before coupling to eliminate residual pump light that would otherwise generate Raman noise during fiber propagation.
The external quantum efficiency of the converter itself (loss from filtering is excluded) is 72.6\%, comprising 89.2\% fiber coupling efficiency, 86.2\% optical element transmission, and 94.5\% internal quantum efficiency. All efficiency values correspond to V-polarized signal light. For H-polarized signal light, slightly different efficiencies are observed, mainly due to polarization-dependent transmission of optical components and interferometer misalignment.

\section{4. Filtering module and polarization analysis}

The filtering module is designed to remove noise primarily originating from the QFC converter, {including the broadband filtering stage and the narrowband filtering stage.}
The filter setup includes a DM, {a bandpass filter (BPF)} at 1522 nm with a 10-nm full width at half maximum (FWHM), an etalon, and a volume Bragg grating (VBG). The detailed experimental setup for the filtering module is shown in Fig. \ref{FigS1}.

{The broadband filtering stage, formed by DM2 and a BPF with 10-nm FWHM, removes residual 1601-nm pump light and non–phase-matched nonlinear noise. Removing the pump is essential because any leaked pump light can directly enter the detection path or generate broadband Raman scattering near 1522 nm when propagating through long fibers. The broadband filter therefore eliminates pump leakage and prevents pump-induced Raman background.}

{To further remove noise close to the converted-photon wavelength, the narrowband filtering stage consists of an etalon with a FWHM of 27 MHz and a free-spectral range (FSR) of 15 GHz, together with a VBG with a FWHM of 25 GHz. The anti-Stokes Raman (ASR) noise produced by the strong pump forms a broad spectral background extending over several terahertz above the pump frequency, which produces noise around the converted-photon wavelength (1522 nm); therefore, narrowband filtering is required at the converted-photon wavelength. The 15-GHz-FSR etalon and 25-GHz-FWHM VBG jointly suppress transmission of neighboring etalon longitudinal modes, resulting in an effective 27-MHz-bandwidth filter centered at 1522 nm.}

The transmission of the filtering module is 66.9\%, including flange connection and fiber polarization controller (91.5\%), etalon (74.9\%), and VBG direction efficiency (97.7\%). 

The polarization analysis module, consists of a motor-HWP, a PBS, and a Faraday rotator, achieving a transmission efficiency of 93\% and coupling efficiency of 82.6\%. The PBS and the Faraday rotator together function as an optical isolator, preventing interference between back-reflected light from the etalon and the input light. {We achieved an overall efficiency of 37.3\% from the converter input port to the fiber connected to SNSPD. The corresponding noise level is $\sim$280 counts per second (CPS), including SNSPD dark counts ($\sim$30 CPS) and PIQFC converter noise ($\sim$250 CPS).}

{We further evaluate the end excitation probability of a write-out photon being finally detected on the SNSPD after 20-km transmission. The excitation probability of a write-out photon already collected into the 780-nm fiber is estimated to be $\sim$3\%. The overall transmission of fiber-fiber connections (Diomand E-2000 connectors) and fiber splicing is 0.80 . The converter overall efficiency is 0.373. 
The SNSPD detection efficiency is 0.88, and the 20-km fiber link provides an additional transmission factor of 0.40. A direct multiplication of all the above contributions yields an expected excitation probability of $\sim$0.315\%, which is higher than the measured value of $\sim$0.205\% on the SNSPD. 
}

{We attribute this discrepancy to the spectral mismatch between the single-photon bandwidth and the narrow passband of the etalon, which allows only part of the photon to be transmitted. Consistent with this physical picture, the temporal wavepacket of the 780-nm write-out photons has a fitted FWHM of 26.43 ns, while the detected telecom photons after frequency conversion and narrowband filtering exhibit a broader temporal wavepacket with a fitted FWHM of 37.25 ns. This temporal broadening reflects the partial transmission of the photon spectrum through the etalon, and this spectral mismatch explains the difference between the nominal excitation probability and the experimentally measured value. We want to emphasize that the current mismatch between the wavepacket and the etalon, which we did not notice as the SNR is quite high for all measurements, can be easily solved by a larger-detuned and wider excitation pulse. A further increase of the etalon transmission from 75\% to $\sim$90\% is planed by (i) using an etalon from a more mature manufacturer to reduce internal losses and ensure better reflectivity matching on both sides, and (ii) slightly relax the etalon bandwidth requirement to improve transmission while maintaining sufficient ASR-noise suppression. }

\section{5. Polarization Compensation Module}
The compensation light ($\sim$10 $\mu$W) with the same frequency of write-out 780-nm photons toggles between the V and D polarization at 10 Hz with the aid of a liquid crystal variable retarder. It is then coupled into the write-out photon fiber and directed into the QFC module via an electrically controlled flip mirror. After frequency conversion and long fiber transmission, the 1522-nm light is directed into a polarimeter via another electrically controlled flip mirror for polarization analysis. An extra home-made shutter is implemented to protect SNSPDs during fiber compensation. Together with a gradient descent algorithm, and a feedback signal sent to the polarization controller, the polarization of the long fiber is automatically compensated. Throughout the whole measurement, polarization compensation is implemented periodically every several minutes to maintain the polarization states of the transmitted photons.

\begin{figure}[ht!]
\centering
\includegraphics[width=8cm]{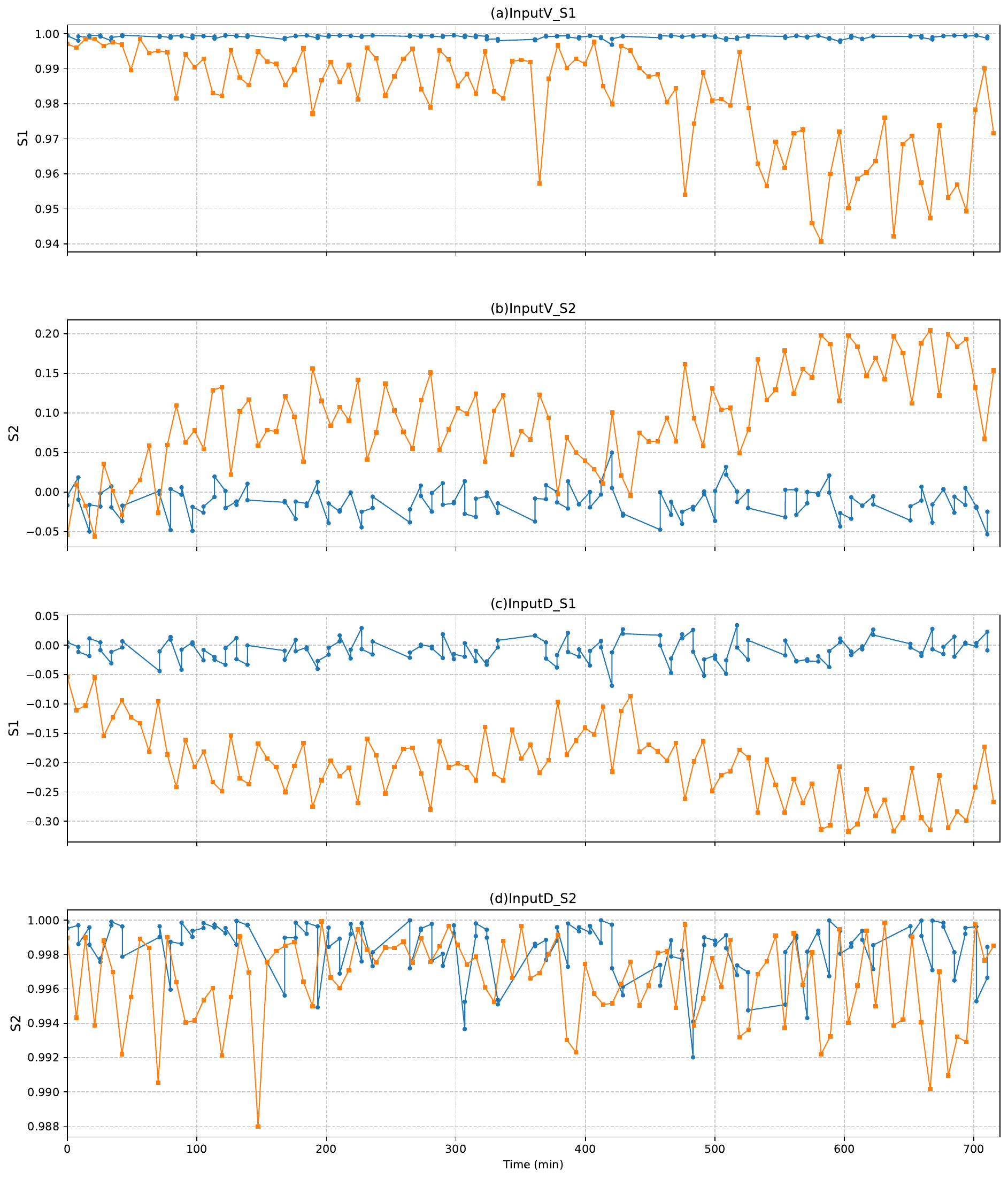}
\caption{ Long-term measurement of the polarization drift after 20-km fiber transmission. The Stokes parameters ${{S}_{1}}$ and ${{S}_{2}}$ of the compensation light are recorded over 12 hours for input V and D polarization. Blue data correspond to measurement with active polarization feedback, while orange data correspond to that without active feedback. The four subplots show: (a) input V state, measured ${{S}_{1}}$; (b) input V state, measured ${{S}_{2}}$; (c) input D state, measured ${{S}_{1}}$; (d) input D state, measured ${{S}_{2}}$.
}
\label{FigS2}
\end{figure}

{To quantify the long-term polarization drift and to demonstrate the performance of the active polarization compensation, we performed a 12-hour measurement of the Stokes parameters ${{S}_{1}}$ and ${{S}_{2}}$ of the compensation light at the output of the 20-km fiber. The input compensation light toggles between the V and D polarization, and the corresponding Stokes parameters were measured using the home-made polarization analysis module described above. The Stokes parameters are defined as
\begin{eqnarray}
{{S}_{1}}=\frac{{{I}_{H}}-{{I}_{V}}}{{{I}_{H}}+{{I}_{V}}},{{S}_{2}}=\frac{{{I}_{A}}-{{I}_{D}}}{{{I}_{A}}+{{I}_{D}}}
\end{eqnarray}
The measured results, shown in Fig. \ref{FigS2}, compare the polarization drift with the active feedback control enabled and disabled. Without active feedback, the Stokes parameters exhibit a slow but continuous drift. When the active feedback control is enabled, the Stokes parameters remain stable for the 12-hour period, confirming the effectiveness of the feedback system in compensating polarization drift for 20-km fiber.}

\section{6. Classical communication system }
{In the main text, we emphasized that the atomic ensemble (in room A) and the QFC module with SNSPDs (in room B) are housed in separate rooms, thus classical communications are necessary for the entanglement verification.
A communicating system including 20-m communicating fibers and three pairs of electro-optic converter and photoelectric converter are used in this experiment for the synchronization of time-critical events.}

{Currently, since we are mainly focused on the entanglement verification, the SNSPDs events, which should be stored locally if in QKD-based communications, are directly send back to room A’s TDC (time to digital converter). For the long fiber compensation, the polarization switching  information of the compensation light (10Hz) is sent from Room A to Room B via one fiber communicating channel. Other signals including fiber compensation start, shutter on/off, or flip mirrors control are done via authenticated network control (SSH connection). The communication system provides a jitter below 200 picoseconds according to an overnight measurement, which is negligible for an experiment with a 100-ns full-width photon wavepacket.}

\section{7. SNR}

The model for simulating the SNR of telecom photons over different fiber lengths is constructed as follows:

\begin{equation}
\begin{split}
\mathrm{SNR}  \propto \frac{(R_{exc}+R_{noise})\times 10^{-\lambda L/10}}{R_{noise}\times 10^{-\lambda L/10}+R_{dark}}\\
\label{Eq.SNR}
\end{split}
\end{equation}
where $R_{exc}, R_{noise}, R_{dark}$ denotes the excitation rate (including repetition rate and excitation efficiency), noise counts from QFC and dark counts, respectively. $L$ is the fiber length, and $\lambda$ represents the attenuation coefficient of the photon in fibers. Using this model, the experimental data are fitted to a fiber length of 100 km with the fitted parameters $ L = 0.2$ dB/km, $R_{noise} = 257$ CPS, $R_{dark} = 38$ CPS.

On the other hand, we also cares about the SNR of read-out photons. Surprisingly, due to the cavity-enhanced scheme, the QFC module introduces additional noise propagating backward to the atomic system, an acousto-optic modulator (AOM) is implemented in the write-out path. This AOM is synchronized to block the noise photons into readout APDs, significantly enhancing the SNR of the read-out photons to roughly at a level of 1500 times the retrieval efficiency. The high SNR of the read-out photons can be easily understood: they constitute a retrieval-efficiency-determined single-photon source.

\bibliography{apscication}